\documentclass[aps, prd, twocolumn, superscriptaddress, nofootinbib, floatfix, noshowpacs]{revtex4-2}

\usepackage{amsmath,amssymb}
\usepackage{graphicx}
\usepackage{grffile}
\usepackage{booktabs,etoolbox}
\usepackage{color,url}
\usepackage[mathscr,scaled=1.15]{urwchancal}
\usepackage{CJKutf8}

\DeclareFontFamily{OT1}{pzc}{}
\DeclareFontShape{OT1}{pzc}{m}{it}{<-> s * [1.15] pzcmi7t}{}
\DeclareMathAlphabet{\mathpzc}{OT1}{pzc}{m}{it}
\definecolor{prdblue}{rgb}{0.133,0.118,0.498}
\usepackage[colorlinks=true,pdfstartview=FitV,linkcolor=prdblue,citecolor=prdblue,urlcolor=prdblue]{hyperref}

\RequirePackage{comment}

\AtBeginEnvironment{thebibliography}{%
	\renewcommand{\href}[2]{#2}%
}

\graphicspath{{Figs/}}

\newcommand{\zeH}{\zeta_{\cal H}}

\newcommand{\GeV}{\mathrm{GeV}}
\newcommand{\dif}{\mathrm{d}}

\begin{document}
\begin{CJK*}{UTF8}{gbsn}

\title{$\,$\\[-6ex]\hspace*{\fill}{\normalsize{\sf\emph{Preprint no}.\ NJU-INP 125-26}}\\[1ex]
Evidence for quark+quark pairing in proton formation}

\author{Hui-Yu Xing (邢惠瑜)%
	$^{\href{https://orcid.org/0000-0002-0719-7526}{\textcolor[rgb]{0.00,1.00,0.00}{\sf ID}}\,}$}
\affiliation{School of Physics, \href{https://ror.org/01rxvg760}{Nanjing University}, Nanjing, Jiangsu 210093, China}
\affiliation{Institute for Nonperturbative Physics, \href{https://ror.org/01rxvg760}{Nanjing University}, Nanjing, Jiangsu 210093, China}
\affiliation{Department of Integrated Sciences and Center for Advanced Studies in Physics, Mathematics and Computation, \href{https://ror.org/03a1kt624}{University of Huelva}, E-21071 Huelva, Spain}

\author{Dan-Dan Cheng（程丹丹)%
    $\,^{\href{https://orcid.org/0009-0009-6466-4483}{\textcolor[rgb]{0.00,1.00,0.00}{\sf ID}}\,}$}
\affiliation{School of Physics, \href{https://ror.org/01rxvg760}{Nanjing University}, Nanjing, Jiangsu 210093, China}
\affiliation{Institute for Nonperturbative Physics, \href{https://ror.org/01rxvg760}{Nanjing University}, Nanjing, Jiangsu 210093, China}

\author{Craig D. Roberts%
	$^{\href{https://orcid.org/0000-0002-2937-1361}{\textcolor[rgb]{0.00,1.00,0.00}{\sf ID}}\,}$}
\email[]{cdroberts@nju.edu.cn}
\affiliation{School of Physics, \href{https://ror.org/01rxvg760}{Nanjing University}, Nanjing, Jiangsu 210093, China}
\affiliation{Institute for Nonperturbative Physics, \href{https://ror.org/01rxvg760}{Nanjing University}, Nanjing, Jiangsu 210093, China}

\author{Jos\'e Rodr\'iguez-Quintero%
       $\,^{\href{https://orcid.org/0000-0002-1651-5717}{\textcolor[rgb]{0.00,1.00,0.00}{\sf ID}}}\,$}
\affiliation{Department of Integrated Sciences and Center for Advanced Studies in Physics, Mathematics and Computation, \href{https://ror.org/03a1kt624}{University of Huelva}, E-21071 Huelva, Spain}

\date{
2026 November 06} 

\begin{abstract}
The question of pairing between quarks inside baryons is addressed by calculating the fragmentation functions that describe production of nucleons (protons and neutrons) in high-energy collisions and comparing attendant predictions with phenomenological global fits to available data and directly with data on hadron multiplicities.  
The fragmentation functions are obtained from nucleon parton distribution functions using the Drell-Levy-Yan relation and subsequently solving a coupled set of pion, kaon, and nucleon jet cascade equations. 
Strong empirical evidence is found for the presence of diquark correlations.  
In comparison with global fits, $d$ (down) quark to proton fragmentation ($d\to p$) exhibits the clearest signals.  
The strongest evidence is found in comparisons with the momentum-dependent proton multiplicity distribution: FFs built from a proton wave function that contains diquark correlations deliver a $z$-profile that is fully consistent with available data, whereas  that obtained from a no-pairing wave function is disfavoured by data at the $9\sigma$ level. 
\\[1ex]
\end{abstract}

\maketitle
\end{CJK*}

\noindent\textbf{1.\ Introduction}\,---\,%
Perceptions of hadron (proton, neutron, pion, etc.) structure have seen a remarkable evolution during the past fifteen years.
New theoretical techniques for study of the strong interaction bound state problem using continuum Schwinger function methods (CSMs) and progress with lattice-regularised quantum chromodynamics (lattice-QCD) have begun to deliver robust insights into the role of quark+quark (diquark)  pairing correlations in building hadrons with $n_{\mathpzc v}\geq 3$ valence degrees of freedom (dofs).

Today, theory suggests that
diquark correlations within $n_{\mathpzc v}=3$ systems (baryons) are a necessary consequence of the dynamics responsible for emergent hadron mass (EHM) \cite{Barabanov:2020jvn, Ding:2022ows, Ferreira:2023fva, Binosi:2026tre} and 
their presence and character make deep impressions on, amongst other things, baryon electroexcitation processes \cite{Ramalho:2023hqd, Achenbach:2025kfx} and helicity dependent baryon structure functions \cite{Yu:2024ovn, Yu:2026poj}.
Experiment has revealed impressions of diquark presence in baryons via the flavour separation of proton elastic electromagnetic form factors \cite{Cates:2011pz, Wojtsekhowski:2026zxn, Cheng:2025yij} and parton distribution functions \cite{JeffersonLabHallATritium:2021usd, Cui:2021gzg, Chang:2022jri}.
Moreover, the potential for diquark correlations to play an important role in $n_{\mathpzc v}= 4, 5$ hadrons (tetraquarks and pentaquarks) is hotly debated \cite{Liu:2019zoy, Guo:2022izw}.  
In this last connection, it is beginning to appear that molecular components of $n_{\mathpzc v} \geq 4$ hadrons dominate except in exceptional circumstances \cite{Bicudo:2022cqi, Hoffer:2024fgm, Francis:2024fwf}.
Such colour-singlet (molecular) subclusters cannot appear in $n_{\mathpzc v} =3$ hadrons, a feature which singles out baryons as, perhaps, the primary residences of diquarks.

Against this background, three key questions stand out, \emph{viz}.\ if diquark correlations are real, then where should physics best look to find, understand, and exploit them? 
New experimental programmes, at existing and anticipated facilities worldwide, \emph{e.g}., see Refs.\,\cite{Zhang:2024lfv, Adams:2018pwt, Aguilar:2019teb, Accardi:2026slw, Shirotori:2015eqa, Aoki:2021cqa, Ai:2025xop}, contain prominent elements that promise to deliver data which may supply answers to these questions.  

The idea that quark+quark pairing might play a role in baryons is as old as the quark model itself \cite{GellMann:1964nj}.  
The relative simplicity of the two-body \emph{cf}.\ three-body problem lent further, practical appeal to the conjecture.
So, pointlike (structureless) diquark components of baryon wave functions were exploited in early model attempts to understand the spectrum and structure of baryons; see, \emph{e.g}., Refs.\,\cite{Ida:1966ev, Lichtenberg:1967zz, Burden:1982ic, Close:1988br}.  

A different picture of diquarks began to appear following the realisation that QCD can support \emph{nonpointlike} colour-charge carrying quark+quark correlations in a variety of isospin-spin combinations \cite{Cahill:1987qr}.  
Notwithstanding the fact that many modellers still elect to work with elementary/pointlike diquark dofs, contemporary continuum and lattice Schwinger function methods provide strong evidence in support of a role in baryon structure for nonpointlike diquark correlations that are essentially nonstatic.
These quark+quark pairs continually break-up, rearrange, and reform within the bound state; see, \emph{e.g}., Refs.\,\cite{Eichmann:2016yit, Burkert:2017djo, Yin:2019bxe, Yin:2021uom, Barabanov:2020jvn, Chen:2021guo, Ding:2022ows, Francis:2025qmy, Gutierrez-Guerrero:2026rsb}.  
This characteristic is driven by many properties of quantum field theory, not least the Pauli exclusion principle for fermions, which requires that in a three valence quark system, no two of those quarks can be singled out for a special role in the wave function.
Naturally, correlations can exist, but all valence quarks must participate equally.

Modern studies suggest that the proton contains both isoscalar-scalar diquarks, $[ud]$, and isovector-axialvector diquarks, $\{ud\}$, $\{uu\}$, with their relative strength being an expression of dynamics; see, \emph{e.g}., Ref.\,\cite[Fig.\,2]{Cheng:2025sdp}.  
(Here, $u = $ up quark and $d = $ down quark.  The neutron is obtained by isospin mapping, \emph{viz}.\ $d^{p,n}\leftrightarrow u^{n,p}$.)
Herein, $s$ (strange) and $c$ (charm) quarks and gluons, $g$, also play a role.  
Notably, nucleon excitations and baryons with heavier quarks may contain diquarks with different quantum numbers and/or quark flavour combinations; see, \emph{e.g}., Refs.\,\cite{Yin:2021uom, Song:2022csw, Torcato:2023ijg, Xie:2024lfo, Yu:2025fer, Cakir:2026fzd}.

In formulating QCD \cite{Fritzsch:1973pi}, it was stressed that the gluon and quark parton fields used to define the Lagrangian cannot be directly observed: it is only colour-neutral hadrons constituted from these dofs that reach detectors.  
The emergence of colour-neutral hadrons from the colour-carrying gluon and quark partons that participate in high-energy annihilation and scattering processes is called hadronisation.  
Herein we will argue that clean signals for the presence of quark+quark correlations in the proton may be found in studies of hadronisation.
We do that by calculating proton fragmentation functions using both 
(\emph{i})  an uncorrelated representation of proton hadronisation, $qqq$ -- {\sf pn},
and 
(\emph{ii}) a quark-pair correlated (diquark) picture of proton structure, $q(qq)$ -- {\sf pc};
and confronting them with available data. 

\smallskip

\noindent\textbf{2.\ Elementary fragmentation processes}\,---\,%
Had\-ro\-ni\-sation is described by fragmentation functions (FFs) \cite[Vol.\,1, Ch.\,19]{ParticleDataGroup:2026mpi}, 
which are themselves of great contemporary interest; see, \emph{e.g}., Refs.\,\cite{Bertone:2017tyb, Gao:2024dbv, Soleymaninia:2020bsq, Moffat:2021dji, AbdulKhalek:2022laj}.
FFs are most cleanly extracted empirically from data on semi-inclusive annihilation (SIA) processes, \emph{viz}.\  $e^+ + e^- \to h + X$, where $h$ is the measured hadron into which the partons coalesce and $X$ represents the remaining, unresolved hadronic debris produced by the electron+positron annihilation. 

Following the approach in Refs.\,\cite{Field:1976ve, Field:1977fa}, the initial stage in a theoretical explanation of hadronisation is determination of the elementary fragmentation function (EFF), $d_{\mathpzc q}^h(z)$, \emph{i.e}., the function which describes the first step in which ${\mathpzc q}$, a given colour-carrying degree of freedom (dof), emits the hadron $h$ with a fraction $z$ of its light-front momentum.  
(For interpretation, one may consider the process in an infinite-momentum frame \cite{Harindranath:1996hq} and identify $z$ with the fraction of the parton's longitudinal momentum transferred to the hadron produced.)

Exploiting crossing symmetry and the Drell-Levy-Yan (DLY) relation \cite{Drell:1969jm, Drell:1969wd, Drell:1969wb, Gribov:1972ri, Gribov:1972rt} at the hadron scale, $\zeta_{\cal H}$, the EFF is given by the ${\mathpzc q}$-in-$h$ parton distribution function (DF):
\begin{equation}
	d_{\mathpzc q}^h(z;\zeta_{\cal H})
	\propto z\,{\mathpzc q}_{h}(1/z;\zeta_{\cal H})\,.
	\label{DLYR}
\end{equation}
Suitably normalised, this DF describes the probability that $\mathpzc q$ carries a fraction $\simeq x$ of the light-front momentum of $h$ when the system is resolved by a probe with energy $\zeta_{\cal H}$.  
At $\zeta_{\cal H}$, all properties of a given hadron are carried by its valence quasiparticle dofs.
The existence of such a scale is guaranteed by the theory of QCD effective charges \cite{Grunberg:1980ja, Grunberg:1982fw} and is leveraged in developing the all-orders (AO) evolution scheme \cite{Yin:2023dbw}, which provides a nonperturbative extension of the DGLAP method \cite{Dokshitzer:1977sg, Gribov:1971zn, Lipatov:1974qm, Altarelli:1977zs} for relating such quantities at different resolving scales. 

To proceed, we assume that, in a given SIA event, the fragmenting dof exhausts itself in producing a cascade of pions ($\pi$), kaons ($K$), and nucleons ($N$ -- protons, $p$, and neutrons, $n$).  
This focuses our analysis and is a fair approximation \cite{Hofmann:1988gy}: empirically, 
non-$\{\pi, K, N\}$ hadrons are $< 15$\% of the total yield and nucleons constitute $\gtrsim 70$\% of the baryons produced.
Moreover, in comparing with data, we focus entirely on the $\{\pi, K, N\}$ yields; so, omitted hadrons do not affect our conclusions.

Working within the framework sketched above and capitalising on $\mathcal G$-parity symmetry, one needs 
three EFFs for pion and kaon hadronisation \cite{Xing:2025eip}
plus, in scenario (\emph{i})\,-\,{\sf pn}, two more, describing $u\to p$, $d\to p$,
or, in scenario (\emph{ii})\,-\,{\sf pc}, five more, \emph{viz}.\ implementations of 
$u\to p$, $d\to p$, $[ud]\to p$, $\{ud\}\to p$, $\{uu\}\to p$.

\begin{table}[t]
	\caption{\label{tab:df_parameters}
		Interpolation coefficients for hadron-scale quark- and diquark-in-proton DFs; see Eq.\,\eqref{eq:proton_df_shape}.}
	\begin{ruledtabular}
		\begin{tabular}{crrrr}
			Eq.\,\eqref{protonDFI}
			& $ u_p$
			& $d_p$
			& $[ud]_p$
			& $\{ud\}_p$
			 \\
			\hline
			${\mathpzc n}$ & $1.155$ & $0.683$ & $1.314$ & $0.140$  \\
			$\rho$         & $1.066$ & $0.973$ & $0.629$ & $0.538$  \\
			\hline
			$a_0$          & $0.040$ & $0.038$ & $0.703$ & $0.702$ \\
			$a_1$          & $5.866$ & $5.686$ & $9.506$ & $9.610$  \\
			$a_2$          & $81.881$ & $79.023$ & $6.330$ & $6.159$  \\
			\hline
			$b_0$          & $0.001$ & $0.001$ & $0.239$ & $0.260$  \\
			$b_1$          & $0.470$ & $0.582$ & $1.245$ & $1.242$ \\
			$b_2$          & $28.919$ & $53.162$ & $4.548$ & $4.298$ \\	
		\end{tabular}
	\end{ruledtabular}
\end{table}

\begin{figure}[t]
\begin{center}
\begin{tabular}{l}
\vspace*{1.5ex}\\
\hspace*{0.39\textwidth}{\large\sf A} \\[-7ex]
\includegraphics[clip, width=0.42\textwidth]{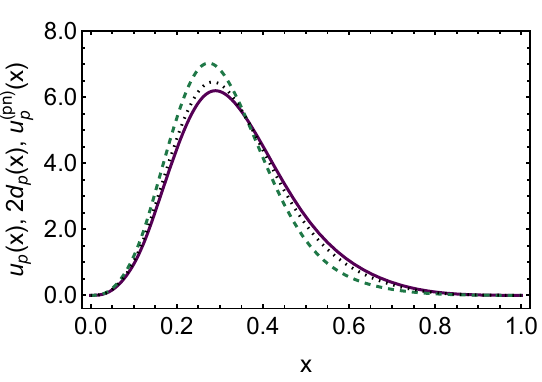} \\
\vspace*{0ex}\\
\hspace*{0.39\textwidth}{\large\sf B} \\[-6ex]
\includegraphics[clip, width=0.42\textwidth]{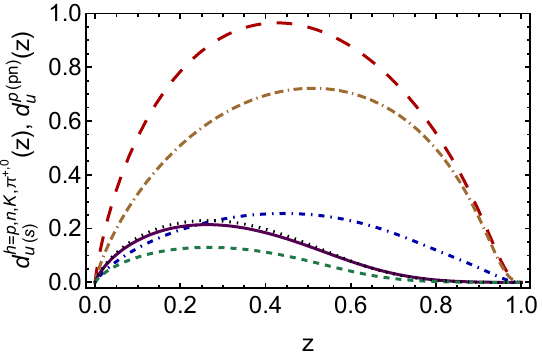} 
\end{tabular}
\end{center}
\vspace*{-4ex}

\caption{\label{proton quark dfs and effs}
{\sf Panel A}. 
In-proton hadron-scale DFs from Ref.\,\cite{Chang:2022jri}:
$ {\mathpzc u}_p(x)$ -- solid purple curve; $2 {\mathpzc d}_p(x)=2 {\mathpzc u}_n(x)$  -- dashed green curve. 
In-proton DFs in scenario (\emph{i})\,-\,{\sf pn}, Eq.\,\eqref{NoPairingDFs}: 
${\mathpzc u}_p^{(\mathsf{pn})}(x) = 2 {\mathpzc d}_p^{(\mathsf{pn})}(x)$ -- dotted black curve. 
{\sf Panel B}. 
Scenario (\emph{i})\,-\,{\sf pn}.
$u \rightarrow p$ EFF: $d_u^{p(\mathsf{pn})}(z) = 2 d_u^{n(\mathsf{pn})}(z)$ -- dotted black curve. 
Scenario (\emph{ii})\,-\,{\sf pc}.
$u \rightarrow p, n$ EFFs: $d_u^ p(z)$ -- solid purple curve; $d_u^n(z)$ -- short-dashed green curve. 
The meson EFFs are common to both frameworks.
$u \rightarrow \pi^{+,0}, K^+$: $d_u^{\pi^{+,0}}(z)$ -- long-dashed red curve; $d_u^ {K^{+}}(z)$ -- dot-dashed blue curve;
$d_s^ {K^{-}}(z)$ -- dash-dash-dotted brown curve.
}
\end{figure}

We take the meson EFFs from Ref.\,\cite{Xing:2025eip}.  The new pieces are those associated with proton hadronisation.  
Consider first $u\to p$, $d\to p$.  Hadron-scale in-proton valence $u$, $d$ DFs were calculated in Ref.\,\cite{Chang:2022jri} within scenario (\emph{ii})\,-\,{\sf pc}.  Reliable interpolations are obtained using 
\begin{subequations}
\label{protonDFI}
\begin{align}
	{\mathpzc q}_{p}(& x;\zeH)
	= {\mathpzc n}
	\ln\left[
	1+\frac{[x^2(1-x)^2]^{3/2}}{\rho^2}
	[1+\mathcal{W}(x)]
	\right] ,
	\label{eq:proton_df_fit} \\
	\mathcal{W}& (x)
	=
	\biggl[a_{0}^{2}
	+\sum_{i=1}^{2}a_{i}^{2} x^{2\alpha_i} \biggr]
	\biggl/
	\biggl[b_{0}^{2}
	+\sum_{i=1}^{2}b_{i}^{2}x^{2 \beta_i}\biggr] , 
	\label{eq:proton_df_shape}
\end{align}
\end{subequations}
with $\alpha_1=2.483$, $\alpha_2=6.516$, $\beta_1=4.802$, $\beta_2=8.917$ and the normalisation, ${\mathpzc n}$, and other coefficients listed in Table~\ref{tab:df_parameters}.   These functions are depicted in Fig.\,\ref{proton quark dfs and effs}A.

{\allowdisplaybreaks
Pairing correlations are absent in scenario (\emph{i}), so Eq.\,\eqref{protonDFI} must be amended.  
A key consequence of diquark correlations in the proton is the prediction $ {\mathpzc u}_p(x; \zeH) \neq 2 {\mathpzc d}_p(x; \zeH)$.
In the absence of such correlations, one instead has SU$(2)$ flavour-spin symmetry and $ {\mathpzc u}_p(x; \zeH) \equiv 2 {\mathpzc d}_p(x; \zeH)$.  We implement this by writing 
\begin{subequations}
\label{NoPairingDFs}
\begin{align}
		{\mathpzc u}^{(\mathsf{pn})}_p(x;\zeH)
		&=\frac{2}{3}[{\mathpzc u}_p(x;\zeH)+{\mathpzc d}_p(x;\zeH)]\,,\\
		{\mathpzc d}^{(\mathsf{pn})}_p(x;\zeH)
		&=\frac{1}{3}[{\mathpzc u}_p(x;\zeH)+{\mathpzc d}_p(x;\zeH)]\,,
\end{align}
\end{subequations}
which preserves the DF sum but realigns the $x$-profiles.
}

In the quark+quark pairing scenario, one must also have EFFs for $\mathcal D\to p$, $\mathcal D= [ud], \{ud\}, \{uu\}$.  
These, too, are specified in Ref.\,\cite{Chang:2022jri}.  
They are reliably interpolated using the functions in Eq.\,\eqref{protonDFI} with the coefficients listed in Table~\ref{tab:df_parameters}: one has
$\{ \mathpzc u \mathpzc u\}_p(x;\zeH) = 2 \{ \mathpzc u \mathpzc d\}_p(x;\zeH)$.
The diquark DFs are depicted in Fig.\,\ref{diquark effs}A.

\begin{figure}[t]
\begin{center}
\begin{tabular}{l}
\vspace*{2ex}\\
\hspace*{0.39\textwidth}{\large\sf A} \\[-4.5ex]
\includegraphics[clip, width=0.42\textwidth]{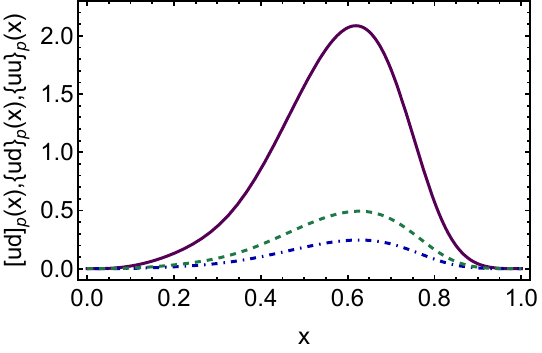} \\
\vspace*{0ex}\\
\hspace*{0.39\textwidth}{\large\sf B} \\[-4.5ex]
\includegraphics[clip, width=0.42\textwidth]{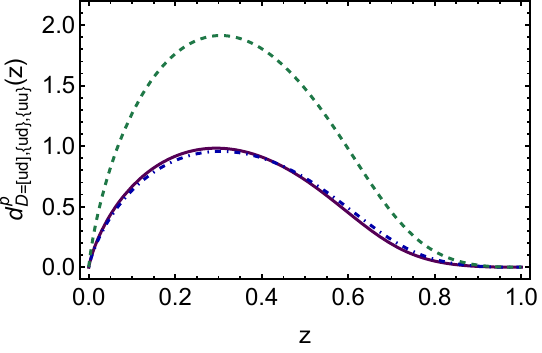}
\end{tabular}
\end{center}
\vspace*{-4ex}

\caption{\label{diquark effs}
{\sf Panel A}. Diquark-in-proton hadron-scale DFs from Ref.\,\cite{Chang:2022jri}. $[ud]_p(x)$ -- solid purple curve;  $\{ud\}_p(x)$ -- dot-dashed blue curve; $\{uu\}_p(x)$ -- dashed green curve. 
{\sf Panel B}. \label{diquark effs}$[ud], \{ud\}, \{uu\}\rightarrow p$ elementary FFs: $d_{[ud]}^{p}(z)$ -- solid purple curve; $d_{\{ud\}}^ {p}(z)$ -- dot-dashed blue curve; $d_{\{uu\}}^ p(z)$ --  dashed green curve.
(The antiparticle EFFs are identical.)}
\end{figure}

{\allowdisplaybreaks
One can now use the DLY relation, Eq.\,\eqref{DLYR}, to obtain all required EFFs.  
Each is normalised by recognising that a given fragmenting dof has unit probability for generating at least and not more than one of the hadrons in the shower, \emph{e.g}.:
\begin{subequations}
\begin{align}
	1 & = \int_0^1 dz \, \left[ \tfrac{3}{2} d_u^{\pi^+}(z;\zeta_{\cal H}) + d_u^{K^+}(z;\zeta_{\cal H})+ d_u^{p+n}(z;\zeta_{\cal H})\right] , \label{EFFlightquarksmesons}
	\\
& = \int_0^1dz\,d_{\{uu\}}^p(z;\zeta_{\cal H}) =\int_0^1dz\,d_{\{dd\}}^n(z;\zeta_{\cal H}) , \\
& = 2 \int_0^1 dz\,d_{[ud]}^p(z;\zeta_{\cal H}) = 2 \int_0^1 dz\,d_{\{ud\}}^p(z;\zeta_{\cal H}) .
\end{align}
\end{subequations}
In Eq.\,\eqref{EFFlightquarksmesons}, the individual integrand terms contribute in the ratio $4$:$1$:$1$ -- the meson fractions were fixed in Ref.\,\cite{Xing:2025eip} 
and the nucleon fraction matches that for kaons because, although the kaon is half as massive as the proton, 
$u$ can only produce $K^+$ but both $p$ and $n$.
The EFFs thus obtained are drawn in Figs.\,\ref{proton quark dfs and effs}B, \ref{diquark effs}B.
}

\begin{figure*}[t]
	\hspace*{-1ex}\begin{tabular}{@{}c@{\hspace{0.05\textwidth}}c@{}}
		\includegraphics[clip, width=0.42\textwidth]{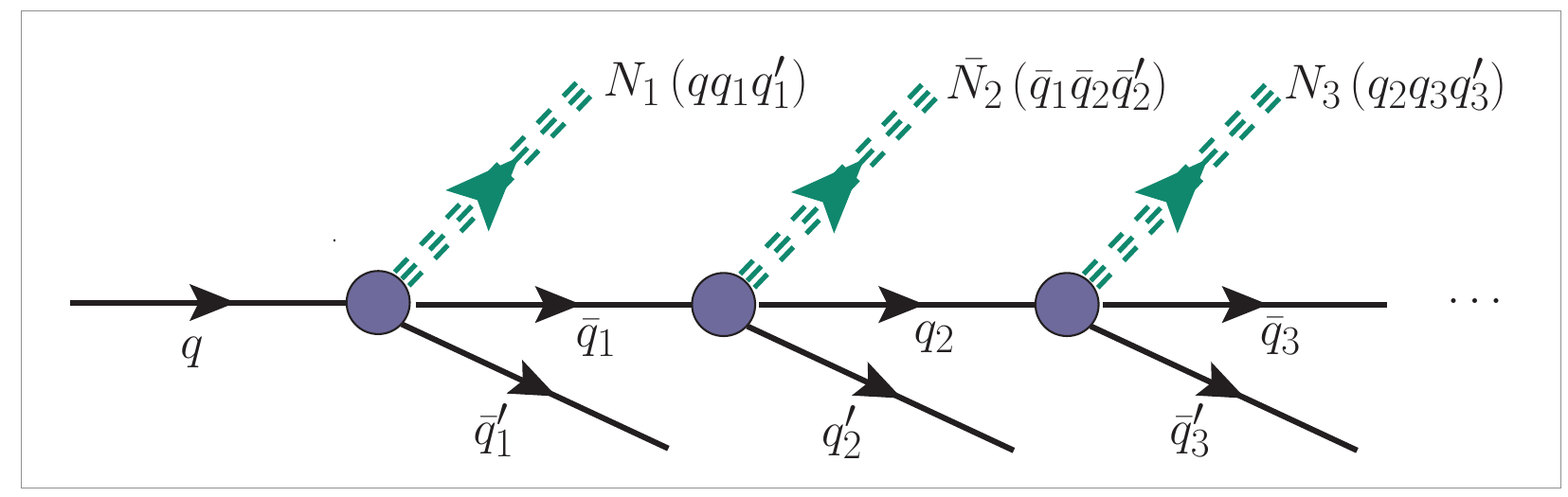}  & 
		\includegraphics[clip, width=0.52\textwidth]{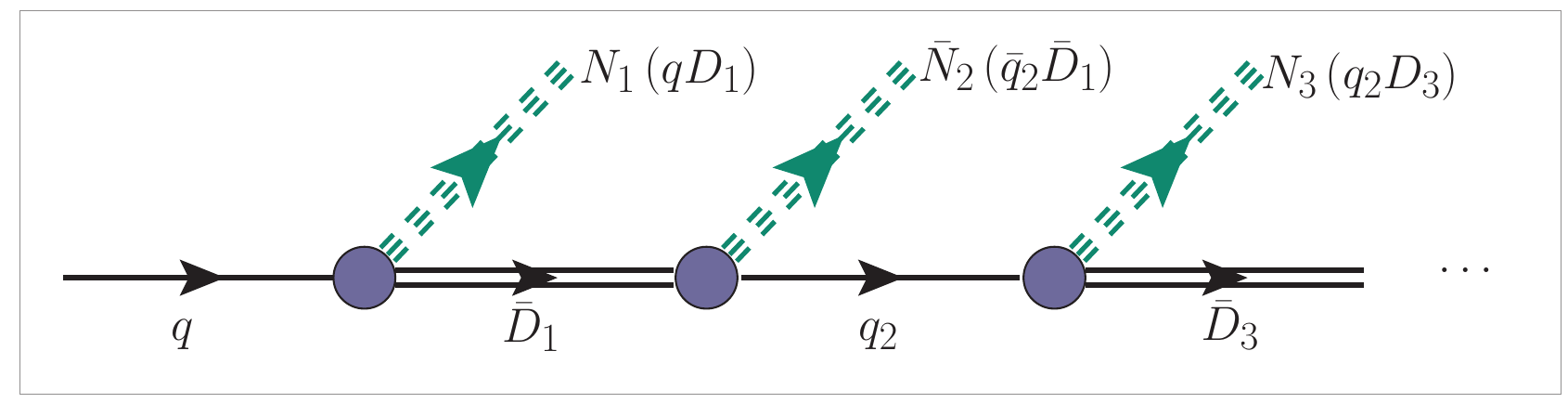} \\[-18ex]
		{\sf A} \hspace*{20em} & {\sf B} \hspace*{25em} \\[14ex]
	\end{tabular}
	\caption{\label{proton diquark dfs and effs}
		Quark-initiated cascades for nucleon production characterising the two distinct schemes considered herein.
		{\sf Panel A}.  Scenario (\emph{i}) -- diquark correlations play no role, with all valence quark dofs fragmenting independently, {\sf pn}.
		{\sf Panel B}.  Scenario (\emph{ii}) -- nucleons emerging from diquark and quark fragmentation, {\sf pc}.
		These cascades are coupled into a tower of coupled jet fragmentation equations.
}
\end{figure*}

Evidently, like DFs, hadron-scale EFFs vanish at the endpoints $z=0,1$;
and light meson EFFs are skewed less toward $z=0$ than those of light diquarks, which are themselves less skewed than light baryon EFFs.
These remarks highlight that elementary processes prefer to deposit a greater fraction of the fragmenting dof's momentum into lighter systems with fewer valence constituents.

\smallskip

\noindent\textbf{3.\ Cascade equations}\,---\,%
Following Ref.\,\cite{Field:1977fa}, the complete hadron-scale FFs, $D_q^h(z;\zeta_{\cal H})$, are obtained by solving a system of coupled jet cascade equations.  
In earlier applications of this sort of approach to proton fragmentation \cite{Sukhatme:1981ym, Bartl:1982bi} it was assumed that diquark-to-meson EFFs are nonzero.  
From the DLY relation perspective, Eq.\,\eqref{DLYR}, this corresponds to supposing that light mesons contain a measurable diquark+antidiquark component.  Today, there is no evidence to support this position; so, we do not include such EFFs.  
On the other hand, we compare hadronisation outcomes in the no-pairing and pairing scenarios, which are distinguished by the series of emissions sketched in Fig.\,\ref{proton diquark dfs and effs}.
For subsequent reference, we note that quark FFs are typically formed into singlet, $S$, and nonsinglet, $N$, combinations: 
\begin{subequations}
	\label{proton singlet and nonsinglet FFs}
	\begin{align}
		D_{S_q}^{h}(z) &=D_q^{h}(z)+D_{\bar q}^{h}(z) , \\
		D_{N_q}^{h}(z) &=D_q^{h}(z)-D_{\bar q}^{h}(z)	=D_q^{h}(z)-D_q^{\bar h}(z).
		\label{proton nonsinglet FF}
	\end{align}
\end{subequations}

Figure~\ref{proton diquark dfs and effs}A illustrates the nucleon piece of the no-pairing hadronisation approach.  
A quark $q$ emits a nucleon $N_1$, leaving two antiquark remnants.  
These remnants fragment separately via independent cascades.
The tower of processes is expressed in the following system of fifteen equations:
\begin{subequations}
	\label{p cascade equations of independent quark model}
	\begin{align}
		\label{p cascade equations of independent quark model 1}
		D_{\mathpzc u}^{h}(z)
		&=d_{\mathpzc u}^{h}(z)
		+\int_z^1\frac{\dif y}{y}
		\sum_{\mathpzc q= \mathpzc u,\mathpzc d, \mathpzc s} d_{\mathpzc u}^{{\mathpzc q}}(z/y) D_{\mathpzc q}^{h}(y)\notag\\
		+\int_z^1&\frac{\dif y}{y}
		\int_{z/y}^1\frac{\dif x}{x}
		\sum_X
		\sum_{\bar q=\bar {\mathpzc u},\bar {\mathpzc d}}
		d_{\mathpzc u}^{X}(z/xy)f_{X}^{\bar {\mathpzc q}}(x)D_{\bar {\mathpzc q}}^{h}(y),\\
		\label{p cascade equations of independent quark model 2}
		D_{\mathpzc s}^{h}(z)
		&=d_{\mathpzc s}^{h}(z)
		+\int_z^1\frac{\dif y}{y}
		\sum_{\mathpzc q = \mathpzc u, \mathpzc d}d_{\mathpzc s}^{\mathpzc q}(z/y)D_{\mathpzc q}^{h}(y).
	\end{align}
\end{subequations}
Here, $h= \pi^{\pm}, K^\pm, K^0, \bar K^0, p, \bar p, n, \bar n$;
$d_{\mathpzc q}^{r}(z)=d_{\mathpzc q}^h(1-z)$, where $r$ is the survivor of the process $q\to h$, 
\emph{e.g.}, $d_{\mathpzc u}^{\mathpzc u}(z)=d_{\mathpzc u}^{\pi^0}(1-z)$;
the $h=n,p$ EFFs are those obtained from Eq.\,\eqref{NoPairingDFs} via the DLY relation; 
$X$ denotes the two antiquark remnants in Fig.\,\ref{proton diquark dfs and effs}A, where
$X_{\bar q_1\bar q_2}\equiv (\bar q_1,\bar q_2)$ specifies the flavour content of this system
and the sum runs over $X_{\bar u\bar d}$, $X_{\bar d\bar d}$, which are the remainders from $u\to p$ and $u\to n$, respectively;
$d_{\mathpzc u}^{X}(\xi)$ is the EFF for $u\to N+X$, with $\xi$ being the two-antiquark remnant fraction of the parent-quark momentum;
and, suitably normalised, 
$f_X^{\bar q}(x)$ is the probability that $\bar q$ carries a fraction $x$ of the total light-front momentum of $X$.
Assuming isospin symmetry, the $f_X^{\bar {\mathpzc q}}(x)$ functions satisfy:
\begin{equation}
		f_{X_{\bar {\mathpzc d}\bar {\mathpzc d}}}^{\bar {\mathpzc d}}(x)
		= 2f_{X_{\bar {\mathpzc u}\bar {\mathpzc d}}}^{\bar {\mathpzc u}}(x)
		=2f_{X_{\bar {\mathpzc u}\bar {\mathpzc d}}}^{\bar {\mathpzc d}}(x) , \quad 
		f_{X_{\bar {\mathpzc d}\bar {\mathpzc d}}}^{\bar {\mathpzc u}}(x) =0.
\end{equation}

We find that proton FFs are insensitive to the profile of $f_X^{\bar q}(x)$.  The results reported herein are obtained using $f_{X_{\bar u\bar d}}^{\bar u}(x)=30x^2(1-x)^2$, which is a DF analogue \cite{Chang:2014lva} of the QCD asymptotic parton distribution amplitude \cite{Lepage:1980fj}. 

{\allowdisplaybreaks
Figure~\ref{proton diquark dfs and effs}B sketches the nucleon part of the quark+quark pairing approach to hadronisation.  
Step 1: a quark, $q$, emits a nucleon, $N_1$, leaving an antidiquark, $\bar D_1$.
Step 2: $\bar D_1$ seeds an antinucleon, $\bar N_2$, leaving a quark, $q_2$, to propagate the cascade.  
These fragmentations continue in tandem with the generation of $\pi$, $K$ mesons, in processes described by a set of thirty-five coupled jet cascade equations:
\begin{subequations}
	\label{cascade equations of diquark model for proton FFs, simple form}
	\begin{align}
		\label{cascade equations of diquark model for proton FFs, simple form1}
		D_{\mathpzc u}^{h}(z)
		&=d_{\mathpzc u}^{h}(z)
		+\int_z^1\frac{\dif y}{y}
		\sum_{{\mathpzc r}\in R}
		d_{\mathpzc u}^{\mathpzc r}(z/y)D_{\mathpzc r}^{h}(y),\\
		D_{\mathpzc s}^{h}(z)
		&=d_{\mathpzc s}^{h}(z)
		+\int_z^1\frac{\dif y}{y}
		\sum_{{\mathpzc q}= {\mathpzc u}, {\mathpzc d}}d_{\mathpzc s}^{\mathpzc q}(z/y)D_{\mathpzc q}^{h}(y) , \label{cascade equations of diquark model for proton FFs, simple form2}\\
		D_{\mathpzc D_1}^h(z)
		& = d_{\mathpzc D_1}^h(z)
		+\int_z^1\frac{\dif y}{y}\sum_{{\mathpzc q} = {\mathpzc u}, {\mathpzc d}}
		d_{\mathpzc D_1}^{{\mathpzc q}}(z/y)D_{\mathpzc q}^h(y), 
		\label{cascade equations of diquark model for pi, k FFs, simple formC}\\
		D_{\mathpzc D_2}^h(z)
		&=d_{\mathpzc D_2}^h(z)
		+\int_z^1\frac{\dif y}{y}\,
		d_{\mathpzc D_2}^{\mathpzc u}(z/y)D_{\mathpzc u}^h(y) . 
		\label{cascade equations of diquark model for pi, k FFs, simple formD}
	\end{align}
\end{subequations}
Here, 
${\mathpzc D_1} = [\bar u \bar d], \{\bar u \bar d\}$, 
${\mathpzc D_2} =\{\bar d \bar d\}$, 
and $R=\{u,d,s,[\overline{ud}],\{\overline{ud}\},\{\overline{dd}\}\}$
labels the remnant states in a $u$-initiated cascade;
the EFFs in Eqs.\,\eqref{cascade equations of diquark model for proton FFs, simple form1}, \eqref{cascade equations of diquark model for proton FFs, simple form2} are those described above in connection with scenario (\emph{ii});
and the antidiquark EFF inputs to Eqs.\,\eqref{cascade equations of diquark model for pi, k FFs, simple formC}, \eqref{cascade equations of diquark model for pi, k FFs, simple formD} are $d_{[\bar{u}\bar{d}]}^{\bar p}$, $d_{\{\bar{u}\bar{d}\}}^{\bar p}$, $d_{\{\bar{u}\bar{u}\}}^{\bar p}$, drawn in Fig.\,\ref{diquark effs}B.
Using $\mathcal G$-parity and charge-conjugation symmetries, all FFs can be obtained from those introduced to and obtained from Eq.\,\eqref{cascade equations of diquark model for proton FFs, simple form}.
}

\smallskip

\noindent\textbf{4.\ Scale evolution}\,---\,%
To enable comparisons with data and phenomenology, it is necessary to evolve the hadron scale FFs to the energy associated with a given empirical analysis.  We employ the AO scheme detailed in Ref.\,\cite[Sec.\,4]{Xing:2025eip}.  
This extension of the standard approach \cite{Dokshitzer:1977sg, Gribov:1971zn, Lipatov:1974qm, Altarelli:1977zs} onto QCD's nonperturbative domain has proven itself in numerous successful applications; see, \emph{e.g}., Refs.\,\cite{Yu:2024ovn, Xu:2024nzp, Wang:2025usl, Xu:2026mds} for recent studies. 

Typically, one need not specify the form of the effective charge associated with the AO approach, $\alpha_{\rm A}$; but if one employs the QCD process-independent (PI) effective charge \cite{Cui:2019dwv}, then $\zeta_{\cal H}=0.331(2)\,\GeV$.
Notably, the PI charge is practically indistinguishable from the process-dependent charge extracted empirically from the Bjorken sum rule \cite{Brodsky:2024zev}, which itself provides one of the most basic constraints on our knowledge of nucleon spin structure.

Herein, the scheme in Ref.\,\cite{{Xing:2025eip}} is augmented by an additional physics input.  
Namely, at $\zeta_{\cal H}$, one has quasiparticles that hadronise to form mesons and baryons.
The quasiparticles contain parton-like objects.  
Only when the energy contained in the fragmentation event exceeds the mass of the hadron to be produced, can the parton-like objects participate in the associated hadronisation process.
Hence, for those hadrons whose mass cannot be neglected, like the proton, the following threshold factor:
\begin{equation}
{\cal P}_h^\zeta  
= \frac{1}{2}\left[1+\tanh\left( \frac{\zeta^2 - \mu_{h{\rm t}}^2}{\zeH^2}\right)\right]
\label{eq:threshold}
\end{equation}
is introduced into the AO FF evolution equations.  Here, $\mu_{h{\rm t}}$ is the active mass threshold; and, for interpretation, note that $P_h^\zeta \approx \theta(\zeta - \mu_{h{\rm t}})$.
We use $\mu_{h{\rm t}} = m_h (1\pm0.20)$, where $m_h$ is the mass of the hadron involved and the $\pm 20$\% serves to supply an estimate of the uncertainty associated with this procedure.

\begin{figure*}[t]
	\hspace*{-1ex}\begin{tabular}{ccc}
		\includegraphics[clip, width=0.31\textwidth]{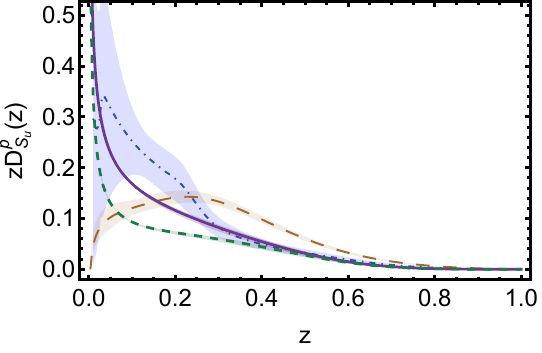} &
		\includegraphics[clip, width=0.31\textwidth]{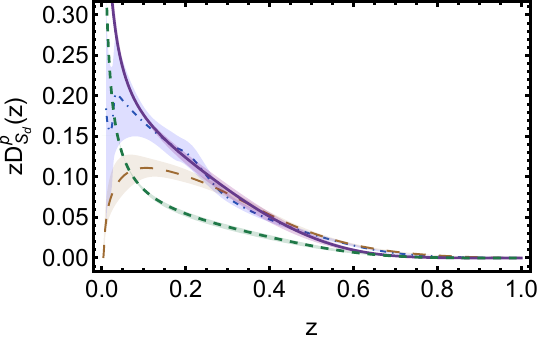} &
		\includegraphics[clip, width=0.31\textwidth]{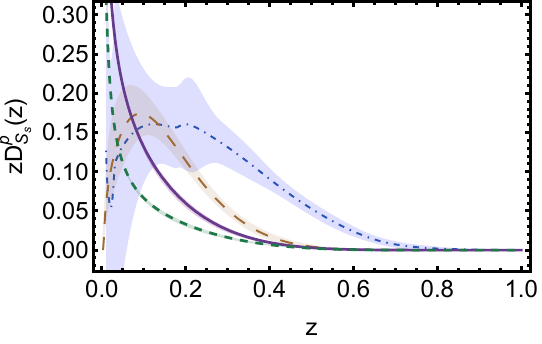} \\[-26ex]
		\hspace*{30ex}{\sf A} & \hspace*{30ex}{\sf B} & \hspace*{30ex}{\sf C} \\
	\end{tabular}\vspace*{22ex}
	\begin{tabular}{ccc}
		\includegraphics[clip, width=0.32\textwidth]{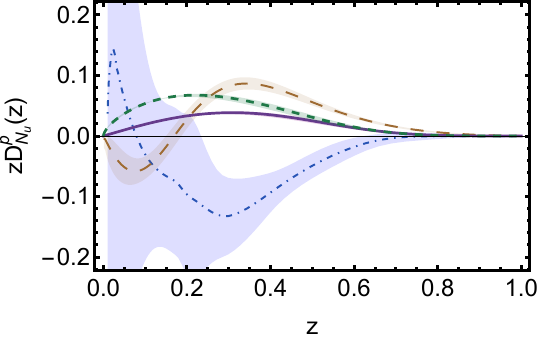} &
		\includegraphics[clip, width=0.32\textwidth]{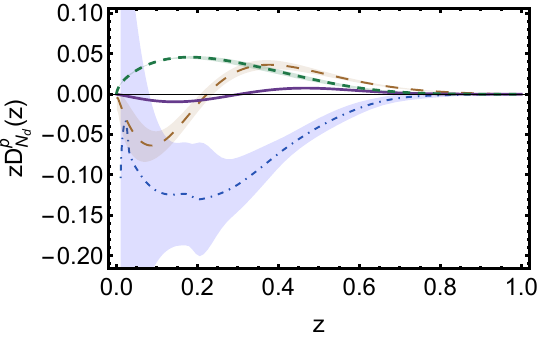} &
		\includegraphics[clip, width=0.32\textwidth]{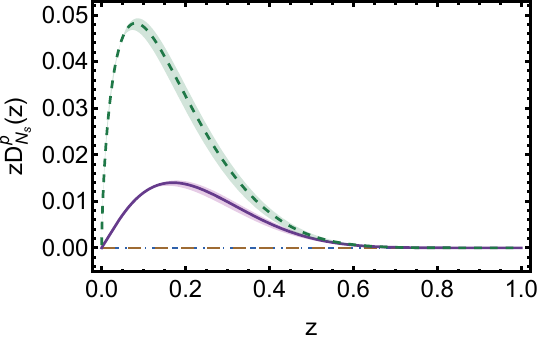}  \\[-26ex]
		\hspace*{30ex}{\sf D} & \hspace*{30ex}{\sf E} & \hspace*{30ex}{\sf F} \\
	\end{tabular}
\vspace*{22ex}
	\caption{\label{proton csm ff results}
	CSM proton FFs obtained from the two scenarios with evolution to $\zeta= 10\,$GeV:
	scenario (\emph{i})\,-\,{\sf pn}  -- dashed green curves within like-coloured bands; 
	scenario (\emph{ii})\,-\,{\sf pc} -- solid purple curves and like-coloured bands.
	The uncertainty bands of CSM predictions are obtained as described  in Sec.\,4.
	Comparison curves: 
	Ref.\,\cite[NNFF]{Bertone:2017tyb} -- dot-dashed blue curves and associated bands;
	Ref.\,\cite[NPC]{Gao:2024dbv} -- long-dashed brown curves within like-coloured bands.
	}
\end{figure*}

{\allowdisplaybreaks
Writing $\breve D(z;t)=zD(z;t)$ and $t=\ln\zeta^2$, with $\zeta$ measured in GeV, the FF evolution equations are:
\begin{subequations}
\begin{align}
\frac{d}{dt} & \breve D_{S_{\mathpzc q}^{{\mathpzc q}^\prime}}^h(z;t) 
= \frac{\alpha_{\rm A}(t)}{2\pi} {\cal P}_h^\zeta \int_z^1\,dy
\bigg[ P_{{\mathpzc q}^\prime {\mathpzc q}}(y)  \nonumber \\
& \qquad \left. 
\times \breve D_{S_{\mathpzc q}^{{\mathpzc q}^\prime}}^h(z/y;t)  + 2 P_{{\mathpzc g}{\mathpzc q}^\prime}(y) \breve D_{{g}_{\mathpzc q}}^h(z/y;t) \right] ,\\
\frac{d}{dt} & \breve D_{g_{\mathpzc q}}^h(z;t) = \frac{\alpha_{\rm A}(t)}{2\pi} {\cal P}_h^\zeta \int_z^1\,dy
\bigg[ \sum_{{\mathpzc q}^\prime} P_{{\mathpzc q}^\prime {\mathpzc g}}(y) \nonumber \\
& \qquad 
\times \breve D_{S_{\mathpzc q}^{{\mathpzc q}^\prime}}^h(z/y;t)
+ P_{{\mathpzc g}{\mathpzc g}} \breve D_{g_{\mathpzc q}}^h(z/y;t) \bigg] ,
\end{align}
\label{eq:evolution}
\end{subequations}
\hspace*{-0.5\parindent}where 
$P_{ij}$ are the leading-order timelike splitting functions \cite[Eq.\,(14)]{Xing:2023pms} and the sum runs over $n_f=4$ active sea quark flavours, $u, d, s, c$.
As in  Ref.\,\cite{Xing:2025eip}, inclusion of a hadron-scale gluon FF ensures momentum conservation under FF evolution.
The evolution equations are solved with the initial conditions:
\begin{equation}
D_{S_{\mathpzc q}^{{\mathpzc q}}}^p(z;\zeta_{\cal H}) = D_{S_{\mathpzc q}}^p(z;\zeta_{\cal H}) ,
\quad 
D_{S_{\mathpzc q}^{{\mathpzc q}^\prime \neq {\mathpzc q}}}^p(z;\zeta_{\cal H}) = 0 ,
\end{equation}
$D_{S_{\mathpzc c}^{{\mathpzc q}}}^p(z;\zeta_{\cal H}) = 0$, etc. 
}

\smallskip

\noindent\textbf{5. Evidence for diquark correlations: global fits}\,---\,%
Figure \ref{proton csm ff results} compares CSM proton FFs calculated using both scenarios in Fig.\,\ref{proton diquark dfs and effs} with results obtained via global fits to available data.  
FFs are only shown for those quarks that are valence dofs in one of the hadrons that can appear in the shower and pion and kaon comparisons are available elsewhere \cite{Xing:2023pms}.  

In Fig.\,\ref{proton csm ff results}, we explicitly include results from Ref.\,\cite[NNFF]{Bertone:2017tyb}, inferred from SIA data, 
and Ref.\,\cite[NPC]{Gao:2024dbv}, which also considers data from semi-inclusive deep inelastic scattering and LHC measurements.
These two sets are representative.  
As previously observed and is again evident in the figure, no two sets of available phenomenologically determined FF sets are mutually consistent overall \cite{Xing:2025eip}, something which adds significantly to the value of analyses like ours.  

The following remarks are worth recording.
(\emph{I}) 
The objective uncertainty on the NNFF results highlights that available data provide very little constraint on $z\lesssim 0.3$, in consequence of which we typically do not stress comparisons between our predictions and fits on this domain. 
(\emph{II}) 
There is no reason why $z D_{S}$, where $D_{S}$ is any singlet FF, should remain finite or even vanish as $z\to 0$.
On the contrary, matching the behaviour of singlet DFs, our analyses predict that all functions $z D_{S}$ should diverge slower than a power-law as $z\to 0$.
(\emph{III}) 
Given their definition, Eq.\,\eqref{proton nonsinglet FF}, nonsinglet FFs should vanish as $z\to 0$ owing to cancellations between those contributions (sea and glue) which might produce a divergence.

Each panel in Fig.\,\ref{proton csm ff results} is now discussed in turn.
\begin{description}
\item[Fig.\,\ref{proton csm ff results}\,A.] \underline{$u\to p$ singlet}.
In this dominant fragmentation channel, both CSM scenarios deliver practically equivalent FFs on $z\gtrsim 0.3$. 
On this domain, they match the NNFF inference and disfavour the NPC fit.  
Equivalence between the two CSM scenarios reveals that this channel is not very sensitive to the presence of diquark correlations. 

\item[Fig.\,\ref{proton csm ff results}\,B.] \underline{$d\to p$ singlet}.
In this strong but subdominant channel, the no-pair and pairing CSM scenarios show marked differences: on $z\gtrsim 0.3$, the diquark paired scheme is clearly favoured by both phenomenological fits.

\item[Fig.\,\ref{proton csm ff results}\,C.] \underline{$s\to p$ singlet}.
In this unfavoured channel, both CSM scenarios predict $D_{S_s}^p(z)\simeq 0$ for $z\gtrsim 0.4$ because protons do not contain valence $s$ quarks.
Indeed, $s$ quarks should populate the in-proton sea; hence, deliver support only on $z\lesssim 0.4$.
This feature is expressed in the NPC fit, but that of NNFF conflicts with this physical expectation.

\item[Fig.\,\ref{proton csm ff results}\,D.] \underline{$u\to p$ nonsinglet}.
Both CSM fragmentation scenarios predict $zD_{N_u}^p(z)>0$ on the entire $z$ domain, with quantitative differences on $z\lesssim 0.4$.
Considering the definition, Eq.\,\eqref{proton nonsinglet FF}, \emph{i.e}., 
$D_{N_{\mathpzc u}}^{p}(z) =D_{\mathpzc u}^{p}(z)-D_{\mathpzc u}^{\bar p}(z)$, 
a nonnegative result is natural for the following reasons.
(\emph{a}) 
$d_{\mathpzc u}^p(z;\zeH)\simeq 2d_{\mathpzc d}^p(z;\zeH)>0$; so, the positive $u\to p$ EFF driving term is larger than that for $d\to p$.
(\emph{b}) 
Whilst $d\to p$ fragmentation leaves $\bar u \bar u$, 
$u\to p$ fragmentation leaves the pair $\bar u \bar d$, which can produce either $\bar p$ or $\bar n$. 
Hence, only part of the $u\to p$ fragments enter the $\bar p$ hadronisation jet; 
so, $D_{\mathpzc u}^{p}(z) > D_{\mathpzc u}^{\bar p}(z) \Rightarrow D_{N_{\mathpzc u}}^{p}(z) > 0$, except at the endpoints.
(No correlation need be assumed for this argument.)

\item[Fig.\,\ref{proton csm ff results}\,E.] \underline{$d\to p$ nonsinglet}.
This channel provides the most noticeable distinction between the unpaired and paired scenarios.
Scenario (\emph{i})\,-\,{\sf pn} delivers a nonnegative FF, 
whereas scenario (\emph{ii})\,-\,{\sf pc} predicts a smaller overall magnitude for $zD_{N_{\mathpzc d}}^p(z)$ and a sign change: it is positive at large $z$, crosses zero, and becomes negative at smaller $z$.
A similar pattern is seen in the NPC fit.  

The scenario (\emph{ii})\,-\,{\sf pc} sign change in $D_{N_{\mathpzc d}}^p(z) = D_{\mathpzc d}^{p}(z)-D_{\mathpzc d}^{\bar p}(z)$ results from competition between valence dof dominated $p$ production and sea driven $\bar p$ production.  
On the valence domain, $z\gtrsim 0.4$, the cascade is dominated by the first emission 
$D_{\mathpzc q}^h(z;\zeH) \simeq d_{\mathpzc q}^h(z;\zeH)$; and since $d_{\mathpzc d}^p(z;\zeH)>0$, one has 
$D_{\mathpzc d}^p(z) > D_{\mathpzc d}^{\bar p}(z) \Rightarrow D_{N_{\mathpzc d}}^p(z)>0$. 

On $z\lesssim 0.4$, fragmentations of sea and glue dofs come to dominate; so, subsequent emissions become important. 
In scenario (\emph{i})\,-\,{\sf pn}, $d\to p$ fragmentation leaves $\bar u\bar u$, which can produce $\bar p$, $\bar n$ in fairly equal measure.
The situation is different in scenario (\emph{ii})\,-\,{\sf pc}: 
$d\to p$ leaves solely the $\{\bar u\bar u\}$ antidiquark; 
that pair can only produce $\bar p$ in the next fragmentation process;
consequently, $\bar p$ is enhanced in the ensuing cascade, driving 
$D_{\mathpzc d}^p(z) < D_{\mathpzc d}^{\bar p}(z) \Rightarrow D_{N_{\mathpzc d}}^p(z) < 0$. 

Evidently, the presence of diquark correlations is the origin of the sign change: they force $d\to p$ to favour $\bar p$ production at lower $z$ values.

The NNFF fit is negative on an even larger $z$ domain than predicted by scenario (\emph{ii})\,-\,{\sf pc}.
This might indicate an overestimate of sea quark contributions to proton fragmentation in NNFF, something which is also apparent in the exaggerated large-$z$ support of $s\to p$ in Fig.\,\ref{proton csm ff results}\,C.

\item[Fig.\,\ref{proton csm ff results}\,F.] \underline{$s\to p$ nonsinglet}. 
Fits usually impose $D_{\mathpzc s}^p(z)=D_{\bar {\mathpzc s}}^p(z)$; in which case the $s$-quark nonsinglet FF vanishes by definition.  This is evident in the absence of NPC and NNFF curves in this image.  

On the other hand, both CSM scenarios predict $D_{N_{\mathpzc s}}^p(z)>0$.
This result is readily understood.
Kaon emission is the dominant elementary process in an $s$-initiated jet. 
After kaon emission, 
an $s$ quark leaves a $u$ or $d$ quark, 
whereas an $\bar s$ quark leaves $\bar u$ or $\bar d$.
As indicated by Figs.\,\ref{proton csm ff results}\,D, E, the combined contributions of the $u$, $d$ channels favour proton production in the $s$-initiated cascade over the $\bar s$-initiated cascade.
Consequently, $D_{\mathpzc s}^p(z)>D_{\bar {\mathpzc s}}^p(z)$.
This prediction can be tested in future global fits by allowing $D_{\mathpzc s}^p(z)$ and $D_{\bar {\mathpzc s}}^p(z)$ to vary independently.
\end{description}

Plainly, in the magnitude of the $d$-quark singlet FF [Fig.\,\ref{proton csm ff results}\,B] and, especially, the sign change in the $d$-quark nonsinglet FF [Fig.\,\ref{proton csm ff results}\,E], one has strong evidence for the presence of diquark correlations in the nucleon.

\smallskip

\noindent\textbf{6. Evidence for diquark correlations: data}\,---\,%
A direct comparison with SIA data is also possible.
For $e^+e^-\to hX$, the multiplicity structure function is \cite{Metz:2016swz,Xing:2025eip}:
\begin{equation}
	F^h(z;\zeta)
	= \frac{1}{\sigma_{\rm tot}}\frac{d\sigma^{e^+e^- \to h X}}{dz} , 
\end{equation}
where, at leading order in perturbative QCD, 
\begin{equation}
	\frac{d\sigma^{e^+e^- \to h X}}{dz}
	= \sum_qe_q^2 D_q^h(z;\zeta)\,,
	\label{differential cross section}
\end{equation}
and $\sigma_{\rm tot} = \sum_{q}e_q^2 =: \sigma_{\rm tot}^{\mbox{\rm\tiny  LO}}$.  
The total $h$-multiplicity is
\begin{equation}
	M^h(\zeta)
	=\int_{z_{\rm min}}^1\dif z\,F^h(z;\zeta),
	\label{total multiplicity}
\end{equation}
where $z_{\rm min} = 2m_h/\zeta$ is the kinematic lower limit.

\begin{table}[t]
	\centering
	\caption{\label{tab:multiplicity}
Fractional $\pi$, $K$ and $N$ multiplicities in $e^+ e^- \to h X$ from Refs.\,\cite{Hofmann:1988gy, ParticleDataGroup:2026mpi} compared with CSM predictions as obtained in the two scenarios.  
Here, $\pi=\pi^++\pi^-+\pi^0$, $K=K^++K^-+K^0+\bar K^0$; and since neutron multiplicity data are unavailable at the tabulated energies, $N=2(p+\bar p)$ for both the experimental data and CSM predictions.
\\
}
	\begin{ruledtabular}
		\begin{tabular}{llll}
			 $10\,\GeV$ & $\pi$ & $K$ & $N$\\
			\hline
			 \cite{Hofmann:1988gy, ParticleDataGroup:2026mpi} & $0.80_{\pm(1)}$ & $0.15_{\mp(1)}$ & $0.044_{\mp(2)}$\\
			  (\emph{i}) no pairing  & $0.81_{\pm(1)}$ & $0.16_{\mp(1)}$ & $0.029_{\mp(1)}$\\
			  (\emph{ii}) diquark pairing  & $0.79_{\pm (1)}$ & $0.16_{\mp (1)}$ & $0.046_{\mp (2)}$\\
			 \hline
			\hline
			$29\,\GeV$\\
			\hline
			 \cite{Hofmann:1988gy, ParticleDataGroup:2026mpi} & $0.79_{\pm(1)}$ & $0.15_{\mp(1)}$ & $0.063_{\mp(5)}$\\
			 (\emph{i}) no pairing & $0.80_{\pm(1)}$ & $0.17_{\mp(1)}$ & $0.029_{\mp(1)}$ \\
			 (\emph{ii}) diquark pairing & $0.78_{\pm (1)}$ & $0.17_{\mp (1)}$ & $0.055_{\mp(1)}$
		\end{tabular}
	\end{ruledtabular}
\end{table}

Using our FF predictions and Eq.\,\eqref{total multiplicity}, one obtains the pion, kaon and nucleon multiplicity fractions listed in Table~\ref{tab:multiplicity}.
At both energies, scenario (\emph{ii})\,-\,{\sf pc} delivers agreement with available experimental results, whereas scenario (\emph{i})\,-\,{\sf pn} underestimates the nucleon yield by approximately a factor of two.

In practice, $\sigma_{\rm tot}$ can only be calculated nonperturbatively and any measured value depends on many things, including experimental setup, detector acceptance, etc.  
So, for comparisons with data on the $z$-dependent multiplicity distribution, we write
\begin{equation}
F^h(z;\zeta) = \frac{1}{{\cal N}}
\sum_{\mathpzc q}e_{\mathpzc q}^2 D_{\mathpzc q}^h(z;\zeta)\,,
\label{FhNorm}
\end{equation}
with ${\cal N}$ chosen in order to ensure a fair match between our predictions and experiment for the \emph{integrated} multiplicity over the central domain $z\in[0.2,0.7]$.  This domain is chosen because, in our view, the experimental uncertainty in existing experimental results is underestimated on the complement of this domain, \emph{viz}., the deep sea and far valence domains.
The value of $1/\mathcal N$ is fixed separately for each scenario:
scenario (\emph{i})\,-\,{\sf pn}, $1/\mathcal N=4.19/\sigma_{\rm tot}^{\rm LO}$;
scenario (\emph{ii})\,-\,{\sf pc}, which is more effective at producing nucleons, $1/\mathcal N=2.68/\sigma_{\rm tot}^{\rm LO}$.

\begin{figure}[t]
%
\centerline{\includegraphics[width=0.93\columnwidth]{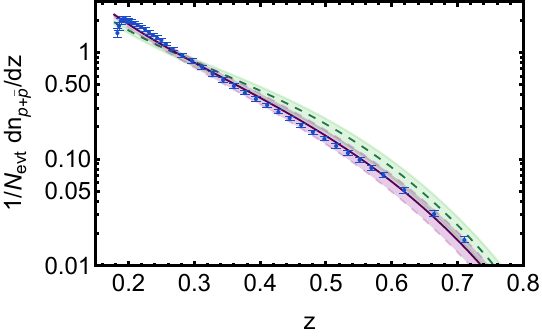}}
\vspace*{-2ex}

\caption{\label{fig:proton_multiplicity}
Hadron multiplicity structure function, Eq.\,\eqref{FhNorm} at $\zeta = 10.5\,$GeV.
Legend.
Scenario (\emph{i})\,-\,{\sf pn} -- dashed green curve within a like-coloured band.
Scenario (\emph{ii})\,-\,{\sf pc} -- solid purple curve within a like-coloured band.
Data from Ref.\,\cite[BaBar]{BaBar:2013yrg}.
}
\end{figure}

The resulting $p+\bar p$ distributions at $\zeta=10.5\,\GeV$ are compared with the BaBar data in Fig.\,\ref{fig:proton_multiplicity}.
For (\emph{i})\,-\,{\sf pn}, $\chi^2_{\rm reduced} = 4.8$, 
whereas (\emph{ii})\,-\,{\sf pc} delivers $\chi^2_{\rm reduced} = 1.2$.
Evidently, the diquark-correlated fragmentation scenario provides an excellent description of the BaBar data, whereas the no-diquark scenario is disfavoured by the data at the $9\sigma$ level. 
Thus, in this comparison, one has additional, very strong support for the presence of diquark correlations in the nucleon.

\smallskip

\noindent\textbf{7. Summary and perspective}\,---\,%
It has long been conjectured that quark+quark pairing plays a key role in forming baryon structure and modern dynamical studies of baryon structure deliver Poincar\'e-covariant wave functions that feature nonpointlike, fully-dynamical diquarks as prominent components. 
The analysis herein confronted these notions and predictions with data on nucleon (proton and neutron) hadronisation as seen in the jets produced by highly energetic collisions.  

Hadronisation is described by parton fragmentation functions (FFs).
We found that a proton wave function which excludes diquark correlations delivers FFs in conflict with global fits to available data. 
On the other hand, predictions obtained using a wave function that contains diquark correlations deliver FFs which agree with the fits on their domain of robust constraint.  The clearest signals for differences between the two pictures are found in proton hadronisation from a $d$ (down) quark.  

Very strong signals for the presence of diquark correlations in the proton were seen in results for hadron multiplicities.  
Wave functions built without quark+quark pairing deliver an integrated nucleon yield that is only half that seen in experiment, whereas predictions obtained using a diquark pairing wave function agree with available data.  
More striking evidence is found in the momentum-dependent multiplicity distribution: the prediction obtained with the diquark paired wave function is fully consistent with available data, whereas that obtained from the unpaired wave function is disfavoured by data at the $9\sigma$ level. 

There are many reasons to extend this study to other members of the baryon octet.  A comparison between $\Lambda$ and $\Sigma^0$ baryons is of special interest because they both have the same valence quark content but, owing to their different isospin quantum numbers, the possible diquark correlations are different in each system.  Based on the analysis herein, one would therefore expect marked differences in their hadronisation patterns, which could be identified/tested in experiment \cite{Bertone:2026fnm}.

\smallskip

\noindent\textit{Acknowledgments}\,---\,%
We acknowledge useful discussions with L.~Chang.
Work supported by:
National Natural Science Foundation of China, grant no.\ 12135007;
and Spanish Ministry of Science and Innovation (MICINN), grant nos.\ PID2022-140440NB-C22, PID2025-171916NB-C32.

\,\\


\begin{thebibliography}{10}
\expandafter\ifx\csname url\endcsname\relax
  \def\url#1{{\tt #1}}\fi
\expandafter\ifx\csname urlprefix\endcsname\relax\def\urlprefix{URL }\fi
\providecommand{\eprint}[2][arXiv]{#1:\linebreak[0]#2}

\bibitem{Barabanov:2020jvn}
Barabanov M~Y {\em et~al\/} 2021 {Diquark Correlations in Hadron Physics:
  Origin, Impact and Evidence} {\em Prog. Part. Nucl. Phys.\/} {\bf 116} 103835

\bibitem{Ding:2022ows}
Ding M, Roberts C~D and Schmidt S~M 2023 {Emergence of Hadron Mass and
  Structure} {\em Particles\/} {\bf 6} 57--120

\bibitem{Ferreira:2023fva}
Ferreira M~N and Papavassiliou J 2023 {Gauge Sector Dynamics in QCD} {\em
  Particles\/} {\bf 6} 312--363

\bibitem{Binosi:2026tre}
Binosi D 2026 {Gauge Symmetry Beyond Perturbation Theory: BRST and anti-BRST
  Structure, Background Fields, and Infrared Dynamics of Yang--Mills Theory}
  {\em Particles\/} {\bf 9} 59

\bibitem{Ramalho:2023hqd}
Ramalho G and Pe{\~n}a M~T 2024 {Electromagnetic transition form factors of
  baryon resonances} {\em Prog. Part. Nucl. Phys.\/} {\bf 136} 104097

\bibitem{Achenbach:2025kfx}
Achenbach P, Carman D~S, Gothe R~W, Joo K, Mokeev V~I and Roberts C~D 2025
  {Electroexcitation of Nucleon Resonances and the Emergence of Hadron Mass}
  {\em Symmetry\/} {\bf 17} 1106

\bibitem{Yu:2024ovn}
Yu Y and Roberts C~D 2024 {Impressions of Parton Distribution Functions} {\em
  Chin. Phys. Lett.\/} {\bf 41} 121202

\bibitem{Yu:2026poj}
Yu Y, Cheng P, Xing H~Y, Binosi D and Roberts C~D 2026 {Helicity Dependent
  Distribution Functions of the proton and {\ensuremath{\Lambda}} and
  {\ensuremath{\Sigma}}0 Baryons} {\em PoS\/} {\bf SPIN2025} 029

\bibitem{Cates:2011pz}
Cates G, de~Jager C, Riordan S and Wojtsekhowski B 2011 {Flavor decomposition
  of the elastic nucleon electromagnetic form factors} {\em Phys. Rev. Lett.\/}
  {\bf 106} 252003

\bibitem{Wojtsekhowski:2026zxn}
Wojtsekhowski B and Cates G~D 2026 {The Super Bigbite Spectrometer Physics
  Program} {\em Particles\/} {\bf 9} 97

\bibitem{Cheng:2025yij}
Cheng P, Yao Z~Q, Binosi D, Lu Y and Roberts C~D 2025 {Quark + diquark
  description of nucleon elastic electromagnetic form factors} {\em Eur. Phys.
  J. A\/} {\bf 61} 255

\bibitem{JeffersonLabHallATritium:2021usd}
Abrams D {\em et~al\/} 2022 {Measurement of the Nucleon $F^n_2/F^p_2$ Structure
  Function Ratio by the Jefferson Lab MARATHON Tritium/Helium-3 Deep Inelastic
  Scattering Experiment} {\em Phys. Rev. Lett.\/} {\bf 128} 132003

\bibitem{Cui:2021gzg}
Cui Z~F, Gao F, Binosi D, Chang L, Roberts C~D and Schmidt S~M 2022 {Valence
  quark ratio in the proton} {\em Chin. Phys. Lett. \emph{Express}\/} {\bf 39}
  041401

\bibitem{Chang:2022jri}
Chang L, Gao F and Roberts C~D 2022 {Parton distributions of light quarks and
  antiquarks in the proton} {\em Phys. Lett. B\/} {\bf 829} 137078

\bibitem{Liu:2019zoy}
Liu Y~R, Chen H~X, Chen W, Liu X and Zhu S~L 2019 {Pentaquark and Tetraquark
  states} {\em Prog. Part. Nucl. Phys.\/} {\bf 107} 237--320

\bibitem{Guo:2022izw}
Guo F~K 2022 {Resolving the mysteries of the positive-parity charm mesons} {\em
  Chin. Sci. Bull.\/} {\bf 67} 4344--4355

\bibitem{Bicudo:2022cqi}
Bicudo P 2023 {Tetraquarks and pentaquarks in lattice QCD with light and heavy
  quarks} {\em Phys. Rept.\/} {\bf 1039} 1--49

\bibitem{Hoffer:2024fgm}
Hoffer J, Eichmann G and Fischer C~S 2025 {Structure of open-flavor four-quark
  states in the charm and bottom region} {\em Phys. Rev. D\/} {\bf 111} 054028

\bibitem{Francis:2024fwf}
Francis A 2025 {Lattice perspectives on doubly heavy tetraquarks} {\em Prog.
  Part. Nucl. Phys.\/} {\bf 140} 104143

\bibitem{Zhang:2024lfv}
Zhang L, Fu J and Cai H 2024 {Advances in the experimental study of pentaquark
  hadrons at LHCb} {\em Chin. Sci. Bull.\/} {\bf 69} 4540--4549

\bibitem{Adams:2018pwt}
Adams B {\em et~al\/} 2018 {Letter of Intent: A New QCD facility at the M2 beam
  line of the CERN SPS (COMPASS++/AMBER) -- arXiv:1808.00848 [hep-ex]$\!$}

\bibitem{Aguilar:2019teb}
Aguilar A~C {\em et~al\/} 2019 {Pion and Kaon Structure at the Electron-Ion
  Collider} {\em Eur. Phys. J. A\/} {\bf 55} 190

\bibitem{Accardi:2026slw}
Accardi A {\em et~al\/} 2026 {Frascati 22 GeV Workshop Summary --
  arXiv:2601.05773 [nucl-ex]} {2nd workshop on Science at the luminosity
  frontier: Jefferson Lab at 22 GEV}

\bibitem{Shirotori:2015eqa}
Shirotori K {\em et~al\/} 2015 {Charmed Baryon Spectroscopy Experiment at
  J-PARC} {\em JPS Conf. Proc.\/} {\bf 8} 022012

\bibitem{Aoki:2021cqa}
Aoki K {\em et~al\/} 2021 {\emph{Extension of the J-PARC Hadron Experimental
  Facility: Third White Paper} -- arXiv:2110.04462 [nucl-ex]$\!$}

\bibitem{Ai:2025xop}
Ai X~C {\em et~al\/} 2025 {Conceptual design report of the Super Tau-Charm
  Facility: the accelerator} {\em Nucl. Sci. Tech.\/} {\bf 36} 242

\bibitem{GellMann:1964nj}
Gell-Mann M 1964 {A Schematic Model of Baryons and Mesons} {\em Phys. Lett.\/}
  {\bf 8} 214--215

\bibitem{Ida:1966ev}
Ida M and Kobayashi R 1966 {Baryon resonances in a quark model} {\em Prog.
  Theor. Phys.\/} {\bf 36} 846

\bibitem{Lichtenberg:1967zz}
Lichtenberg D~B and Tassie L~J 1967 {Baryon Mass Splitting in a Boson-Fermion
  Model} {\em Phys. Rev.\/} {\bf 155} 1601--1606

\bibitem{Burden:1982ic}
Burden C~J and Tassie L~J 1982 {Quarks and Diquarks in a Baryon String Model}
  {\em Nucl. Phys. B\/} {\bf 204} 204--212

\bibitem{Close:1988br}
Close F~E and Thomas A~W 1988 {The Spin and Flavor Dependence of Parton
  Distribution Functions} {\em Phys. Lett. B\/} {\bf 212} 227

\bibitem{Cahill:1987qr}
Cahill R~T, Roberts C~D and Praschifka J 1987 {Calculation of diquark masses in
  QCD} {\em Phys. Rev. D\/} {\bf 36} 2804

\bibitem{Eichmann:2016yit}
Eichmann G, Sanchis-Alepuz H, Williams R, Alkofer R and Fischer C~S 2016
  {Baryons as relativistic three-quark bound states} {\em Prog. Part. Nucl.
  Phys.\/} {\bf 91} 1--100

\bibitem{Burkert:2017djo}
Burkert V~D and Roberts C~D 2019 {Roper resonance: Toward a solution to the
  fifty-year puzzle} {\em Rev. Mod. Phys.\/} {\bf 91} 011003

\bibitem{Yin:2019bxe}
Yin P~L, Chen C, Krein G, Roberts C~D, Segovia J and Xu S~S 2019 {Masses of
  ground-state mesons and baryons, including those with heavy quarks} {\em
  Phys. Rev. D\/} {\bf 100} 034008

\bibitem{Yin:2021uom}
Yin P~L, Cui Z~F, Roberts C~D and Segovia J 2021 {Masses of positive- and
  negative-parity hadron ground-states, including those with heavy quarks} {\em
  Eur. Phys. J. C\/} {\bf 81} 327

\bibitem{Chen:2021guo}
Chen C, Fischer C~S, Roberts C~D and Segovia J 2022 {Nucleon axial-vector and
  pseudoscalar form factors and PCAC relations} {\em Phys. Rev. D\/} {\bf 105}
  094022

\bibitem{Francis:2025qmy}
Francis A 2026 {Updates on diquarks from lattice QCD} {\em PoS\/} {\bf
  HADRON2025} 056

\bibitem{Gutierrez-Guerrero:2026rsb}
Guti\'{e}rrez-Guerrero L~X and Hern\'{a}ndez-Pinto R~J 2026 Symmetry-preserving
  contact interaction approaches: An overview of meson and diquark form factors
  {\em Particles\/} {\bf 9} 45

\bibitem{Cheng:2025sdp}
Cheng P, Liu L, Lu Y and Roberts C~D 2026 {Insights into Nucleon Resonances via
  Continuum Schwinger Function Methods} {\em Acta Phys. Polon. B\/} {\bf 57}
  2--A13

\bibitem{Song:2022csw}
Song Y, Jia D, Zhang W and Hosaka A 2023 {Low-lying doubly heavy baryons: Regge
  relation and mass scaling} {\em Eur. Phys. J. C\/} {\bf 83} 1

\bibitem{Torcato:2023ijg}
Torcato A, Arriaga A, Eichmann G and Pe\~na M~T 2023 {Heavy Baryon Spectroscopy
  in a Quark\textendash{}Diquark Approach} {\em Few Body Syst.\/} {\bf 64} 45

\bibitem{Xie:2024lfo}
Xie J~Q, Song H and Chen J~K 2024 {Regge trajectories for the triply heavy
  bottom-charm baryons in the diquark picture} {\em Eur. Phys. J. C\/} {\bf 84}
  1048

\bibitem{Yu:2025fer}
Yu Y, Cheng P, Xing H~Y, Binosi D and Roberts C~D 2025 {Distribution Functions
  of $\Lambda$ and $\Sigma^0$ Baryons} {\em Eur. Phys. J. A\/} {\bf 61} 208

\bibitem{Cakir:2026fzd}
{\c{C}}ak{\i}r {\"O} and Mutuk H 2026 {Mass spectrum, magnetic moments and
  Regge trajectories of $\Omega _{ccb}$ and $\Omega _{cbb}$ baryons in the
  nonrelativistic quark{\textendash}diquark model} {\em Eur. Phys. J. C\/} {\bf
  86} 678

\bibitem{Fritzsch:1973pi}
Fritzsch H, Gell-Mann M and Leutwyler H 1973 {Advantages of the Color Octet
  Gluon Picture} {\em Phys. Lett. B\/} {\bf 47} 365--368

\bibitem{ParticleDataGroup:2026mpi}
Takahashi F {\em et~al\/} (Particle Data Group) 2026 {Review of Particle
  Physics} {\em Int. J. Mod. Phys. A\/} {\bf 41} 2630011

\bibitem{Bertone:2017tyb}
Bertone V, Carrazza S, Hartland N~P, Nocera E~R and Rojo J 2017 {A
  determination of the fragmentation functions of pions, kaons, and protons
  with faithful uncertainties} {\em Eur. Phys. J. C\/} {\bf 77} 516

\bibitem{Gao:2024dbv}
Gao J, Liu C, Shen X, Xing H and Zhao Y 2024 {Global analysis of fragmentation
  functions to charged hadrons with high-precision data from the LHC} {\em
  Phys. Rev. D\/} {\bf 110} 114019

\bibitem{Soleymaninia:2020bsq}
Soleymaninia M, Goharipour M, Khanpour H and Spiesberger H 2021 {Simultaneous
  extraction of fragmentation functions of light charged hadrons with mass
  corrections} {\em Phys. Rev. D\/} {\bf 103} 054045

\bibitem{Moffat:2021dji}
Moffat E, Melnitchouk W, Rogers T~C and Sato N 2021 {Simultaneous Monte~Carlo
  analysis of parton densities and fragmentation functions} {\em Phys. Rev.
  D\/} {\bf 104} 016015

\bibitem{AbdulKhalek:2022laj}
Abdul~Khalek R, Bertone V, Khoudli A and Nocera E~R 2022 {Pion and kaon
  fragmentation functions at next-to-next-to-leading order} {\em Phys. Lett.
  B\/} {\bf 834} 137456

\bibitem{Field:1976ve}
Field R~D and Feynman R~P 1977 {Quark Elastic Scattering as a Source of High
  Transverse Momentum Mesons} {\em Phys. Rev. D\/} {\bf 15} 2590--2616

\bibitem{Field:1977fa}
Field R~D and Feynman R~P 1978 {A Parametrization of the Properties of Quark
  Jets} {\em Nucl. Phys. B\/} {\bf 136} 1--76

\bibitem{Harindranath:1996hq}
Harindranath A 1996 {An Introduction to light front dynamics for pedestrians --
  hep-ph/9612244} {\em {International School on Light-Front Quantization and
  Non-Perturbative QCD}\/}

\bibitem{Drell:1969jm}
Drell S~D, Levy D~J and Yan T~M 1969 {A Theory of Deep Inelastic Lepton-Nucleon
  Scattering and Lepton Pair Annihilation Processes. 1.} {\em Phys. Rev.\/}
  {\bf 187} 2159--2171

\bibitem{Drell:1969wd}
Drell S~D, Levy D~J and Yan T~M 1970 {A Theory of Deep Inelastic Lepton Nucleon
  Scattering and Lepton Pair Annihilation Processes. 2. Deep Inelastic electron
  Scattering} {\em Phys. Rev. D\/} {\bf 1} 1035--1068

\bibitem{Drell:1969wb}
Drell S~D, Levy D~J and Yan T~M 1970 {A Theory of Deep Inelastic Lepton-Nucleon
  Scattering and Lepton Pair Annihilation Processes. 3. Deep Inelastic
  electron-Positron Annihilation} {\em Phys. Rev. D\/} {\bf 1} 1617--1639

\bibitem{Gribov:1972ri}
Gribov V and Lipatov L 1972 {Deep inelastic $e p$ scattering in perturbation
  theory} {\em Sov. J. Nucl. Phys.\/} {\bf 15} 438--450

\bibitem{Gribov:1972rt}
Gribov V~N and Lipatov L~N 1972 {$e^+ e^-$ pair annihilation and deep inelastic
  $e p$ scattering in perturbation theory} {\em Sov. J. Nucl. Phys.\/} {\bf 15}
  675--684

\bibitem{Grunberg:1980ja}
Grunberg G 1980 {Renormalization Group Improved Perturbative QCD} {\em Phys.
  Lett. B\/} {\bf 95} 70 [Erratum: Phys. Lett. B 110, 501 (1982)]

\bibitem{Grunberg:1982fw}
Grunberg G 1984 {Renormalization Scheme Independent QCD and QED: The Method of
  Effective Charges} {\em Phys. Rev. D\/} {\bf 29} 2315

\bibitem{Yin:2023dbw}
Yin P~L, Xu Y~Z, Cui Z~F, Roberts C~D and Rodr\'\i{}guez-Quintero J 2023
  {All-Orders Evolution of Parton Distributions: Principle, Practice, and
  Predictions} {\em Chin. Phys. Lett. \emph{Express}\/} {\bf 40} 091201

\bibitem{Dokshitzer:1977sg}
Dokshitzer Y~L 1977 Calculation of the structure functions for deep inelastic
  scattering and $e^+$ $e^-$ annihilation by perturbation theory in quantum
  chromodynamics. ({\mbox {i}n {r}ussian}) {\em Sov. Phys. JETP\/} {\bf 46}
  641--653

\bibitem{Gribov:1971zn}
Gribov V~N and Lipatov L~N 1971 {Deep inelastic electron scattering in
  perturbation theory} {\em Phys. Lett. B\/} {\bf 37} 78--80

\bibitem{Lipatov:1974qm}
Lipatov L~N 1975 {The parton model and perturbation theory} {\em Sov. J. Nucl.
  Phys.\/} {\bf 20} 94--102

\bibitem{Altarelli:1977zs}
Altarelli G and Parisi G 1977 {Asymptotic Freedom in Parton Language} {\em
  Nucl. Phys. B\/} {\bf 126} 298--318

\bibitem{Hofmann:1988gy}
Hofmann W 1988 {Particle Composition in Hadronic Jets in $e^+ e^-$
  Annihilation} {\em Ann. Rev. Nucl. Part. Sci.\/} {\bf 38} 279--322

\bibitem{Xing:2025eip}
Xing H~Y, Bian W~H, Cui Z~F and Roberts C~D 2025 {Kaon and pion fragmentation
  functions} {\em Eur. Phys. J. C\/} {\bf 85} 1305

\bibitem{Sukhatme:1981ym}
Sukhatme U~P, Lassila K~E and Orava R 1982 {Diquark Fragmentation} {\em Phys.
  Rev. D\/} {\bf 25} 2975

\bibitem{Bartl:1982bi}
Bartl A, Fraas H and Majerotto W 1982 {Quark and Diquark Fragmentation Into
  Mesons and Baryons} {\em Phys. Rev. D\/} {\bf 26} 1061

\bibitem{Chang:2014lva}
Chang L, Mezrag C, Moutarde H, Roberts C~D, Rodr{\'{\i}}guez-Quintero J and
  Tandy P~C 2014 {Basic features of the pion valence-quark distribution
  function} {\em Phys. Lett. B\/} {\bf 737} 23--29

\bibitem{Lepage:1980fj}
Lepage G~P and Brodsky S~J 1980 {Exclusive Processes in Perturbative Quantum
  Chromodynamics} {\em Phys. Rev. D\/} {\bf 22} 2157--2198

\bibitem{Xu:2024nzp}
Xu Z~N, Binosi D, Chen C, Raya K, Roberts C~D and Rodr\'\i{}guez-Quintero J
  2025 {Kaon distribution functions from empirical information} {\em Phys.
  Lett. B\/} {\bf 865} 139451

\bibitem{Wang:2025usl}
Wang X, Chang L, Ding M, Raya K and Roberts C~D 2026 {Symmetry Constraints on
  Pion Valence Structure} {\em Chin. Phys. Lett.\/} {\bf 43} 070203

\bibitem{Xu:2026mds}
Xu Z~N, Binosi D, Roberts C~D and Rodr\'{\i}guez-Quintero J 2026 {Quest for an
  Understanding of Pion and Kaon Structure - arXiv:2608.07670 [hep-ph]}

\bibitem{Cui:2019dwv}
Cui Z~F, Zhang J~L, Binosi D, de~Soto F, Mezrag C, Papavassiliou J, Roberts
  C~D, Rodr{\'{\i}}guez-Quintero J, Segovia J and Zafeiropoulos S 2020
  {Effective charge from lattice QCD} {\em Chin. Phys. C\/} {\bf 44} 083102

\bibitem{Brodsky:2024zev}
Brodsky S~J, Deur A and Roberts C~D 2024 {The Secret to the Strongest Force in
  the Universe} {\em Sci. Am.\/} {\bf 5 (May)} 32--39

\bibitem{Xing:2023pms}
Xing H~Y, Yao Z~Q, Li B~L, Binosi D, Cui Z~F and Roberts C~D 2024 {Developing
  predictions for pion fragmentation functions} {\em Eur. Phys. J. C\/} {\bf
  84} 82

\bibitem{Metz:2016swz}
Metz A and Vossen A 2016 {Parton Fragmentation Functions} {\em Prog. Part.
  Nucl. Phys.\/} {\bf 91} 136--202

\bibitem{BaBar:2013yrg}
Lees J~P {\em et~al\/} (BaBar) 2013 {Production of charged pions, kaons, and
  protons in $e^+e^-$ annihilations into hadrons at $\sqrt{s}$=10.54 GeV} {\em
  Phys. Rev. D\/} {\bf 88} 032011 (\textit{Preprint} \eprint{1306.2895})

\bibitem{Bertone:2026fnm}
Bertone V, Bongallino A, Chiefa A, Echevarria M~G and Schnell G 2026 {First
  next-to-next-to-leading-order extraction of fragmentation functions for
  {\ensuremath{\Lambda}} hyperons} {\em JHEP\/} {\bf 10} 089

\end{thebibliography}

\providecommand{\newblock}{}

\end{document}